\documentclass[a4paper,10pt]{article}

\usepackage{graphicx}
 \usepackage{epsfig}
\usepackage{graphicx}
 \usepackage{epsfig}
\usepackage{amsmath}
\usepackage{amsfonts}
\usepackage{amssymb}

\title {A van Hemmen-Kondo model for disordered strongly correlated electron systems.}

\author{S. G. Magalhaes$^a$, F. M. Zimmer$^a$, B. Coqblin$^b$
\\
\\
$^a${\it Departamento de F\'isica, Universidade Federal de Santa Maria,}
\\
 {\it Santa Maria 97105-900, RS, Brazil}
\\
$^b${\it L.P.S., CNRS UMR 8502, Universit\'e  Paris-Sud,}
\\{\it 91405-Orsay, France}
}
\date{}
\begin{document}
\maketitle

\begin{abstract}
We present here a theoretical model in order to describe the
competition between the Kondo effect and the spin glass behavior.
The spin glass part of the starting Hamiltonian contains Ising spins
with an intersite exchange interaction given by the local van Hemmen
model, while the Kondo effect is described as usual by the intrasite
exchange $J_K$. We obtain, for large $J_K$ values, a Kondo phase
and, for smaller $J_K$ values, a succession, with decreasing
temperature, of a spin glass phase, a mixed spin glass-ferromagnetic
one and finally a ferromagnetic phase. This model improves the
theoretical description of disordered Kondo systems with respect to
previous models and can account for experimental data in Cerium
disordered systems like $CeCu_{1-x}Ni_x$ alloys.
\end{abstract}

\section{Introduction}

The interplay between disorder and strong electronic correlations is
recognized as a very interesting issue in condensed matter physics.
There are now many experimental evidences showing the very important
role of the disorder in $f$-electron systems in addition to the RKKY
or Kondo interactions \cite{Coqblin}. As a result, it can appear
complex phase diagrams which show spin glass (SG) phases in addition
to the onset of antiferromagnetism (AF) or ferromagnetism (FE),
regions dominated by the Kondo effect, the presence of Quantum
Phase Transitions (QPT) and exotic regions which present non-Fermi
liquid behavior (NFL) \cite{NFL}.

Earlier experimental results can illustrate  the mentioned
complexity. For instance, in $CeAu_{1-x}Co_{x}Si_{3}$ alloys
\cite{Sampa}, when $Au$ is replaced by $Co$, it first appears a SG
phase, then there is the onset of an AF phase with the N\'eel
temperature decreasing towards a Quantum Critical Point (QCP). Thus,
the glassy behaviour tends to decrease with the increase of $x$ and
finally, for $x>0.9$, there is a complete screening of magnetic
moments due to the Kondo effect.

More recently, experimental findings in $CePd_{1-x}Rh_{x}$
\cite{Wester1,Wester2} and $CeNi_{1-x}Cu_{x}$
\cite{Marcano1,Marcano2} have enlarged the set of non-trivial
behaviour in disordered $f$-electron systems. In both systems, there are
strong indications that a glassy behaviour is present in a suitable
range of doping and this behaviour has been recently identified to
appear as a cluster glass state. In the well studied
$CeNi_{1-x}Cu_{x}$ case, the Kondo interaction is dominating for $x$
smaller than approximately 0.2 \cite{Garcia}. However, the
intermediate doping regime has been extensively studied both
experimentally and theoretically and finally a complex scenario is
obtained when the temperature is decreased. In the first
experimental studies on $CeNi_{1-x}Cu_{x}$ alloys with $x$ typically
between 0.3 and 0.6, a SG phase has been obtained below the
paramagnetic state and then there is a transition to a ferromagnetic
phase at lower temperatures.

More sophisticated experiments have recently shown that dynamic
magnetic clusters are developping at low temperatures below the
paramagnetic state. More precisely, there is the formation of
clusters due to short range ferromagnetic correlations below a
certain temperature $T^{*}$. The volume fraction of these clusters
increases as temperature is lowered and they become frozen at
$T_{cl}$ well below $T^{*}$ and, therefore, it appears an
inhomogeneous ferromagnetic order at very low temperatures
\cite{Marcano1,Marcano2}. Thus, there is a change, below the
paramagnetic phase, from a cluster spin glass to a disordered
ferromagnetic order without any sharp transition, but with a mixed
and disordered intermediate phase.

A Kondo-Cluster-Glass state has been also recently evidenced in
$CePd_{1-x}Rh_{x}$ alloys at very low temperatures. This system
exhibits a continuous evolution from a ferromagnetic order in
$CePd$, with a Curie temperature $T_c=6.6 K$, to an
intermediate-valence ground state in $CeRh$. The Curie temperature
decreases continuously with increasing $x$ and tends to 25 mK at the
value $x=0.87$. Despite pronounced non-Fermi-liquid behavior in the
proximity of this concentration for specific heat and thermal
expansion, it was concluded from the analysis of the Gruneisen ratio
that there is no QCP \cite{Pikul}. On the
opposite, a ``Kondo-cluster-glass" state was found for $x$ larger than
0.65: there is firstly the formation of clusters with predominantly
ferromagnetic couplings of the $f$-moments below a given temperature
$T^{*}$ and then a random freezing of the cluster moments below a
smaller temperature $T_{cl}$ \cite{Wester1,Wester2}. Thus, there are
clearly similarities between the low temperature behaviors of
$CeNi_{1-x}Cu_{x}$ and $CePd_{1-x}Rh_{x}$ alloys, but both a more
profound analysis of the different data and the role of the Kondo
effect have to be precised in these two systems.

Several theoretical studies have tried, since already some time, to
account for the previous experimental data. A Kondo lattice with an
additional Ising term and a random coupling between localized
spins, called here the Kondo-Ising Lattice (KIL) model
\cite{Alba1,Magal1,MagalAF1}, has been firstly used to study the
competition between the Kondo effect and magnetism when disorder is
present within the Static Approximation (SA) \cite{Moore}. It
appears that, for $CeAu_{1-x}Co_{x}Si_{3}$ alloys, a Gaussian random
distributed bond would be adequate as can be seen in Refs.
\cite{MagalAF1,MagalAF2}. The same model has been also firstly
used to describe the case of $CeNi_{1-x}Cu_{x}$ alloys, where the
disorder has been introduced within the classical
Sherrington-Kirkpatrick (SK) model \cite{SK} by taking Gaussian
random intersite coupling $J_{ij}$ with a mean value $J_{0}$
different from zero to describe the ferromagnetic  ordering
\cite{Magal1}. The phase diagram giving the temperature $T$ {\it
versus} the strength $J_{K}$ of the Kondo interaction has been
computed and we have obtained, besides the Kondo state, magnetic
phases like Spin Glass (SG), Ferromagnetic (FE) and a mixed phase
(SG+FE). For this particular solution, the ferromagnetic order
occurs with replica symmetry breaking. This phase diagram
could be, therefore, a good starting point to describe the scenario found in
$CeNi_{1-x}Cu_{x}$ alloys. Unfortunately, for this particular kind
of disorder, the Curie temperature $T_{c}$ is always higher than the
freezing one, which is a scenario opposite to the experimental
situation observed in $CeNi_{1-x}Cu_{x}$ system.

Thus, in order to solve the preceding difficulty, a completely
different perspective has been adopted in reference \cite{Magal2}.
The theoretical description of the disorder has been modified from a
bond disordered coupling to a site disordered one. In that case, the
$J_{ij}$ coupling is a generalization of the Mattis model
\cite{Mattis} used extensively to study complex systems \cite{Amit},
given as
$J_{i_{}j_{}}=\frac{J}{2N}\sum_{\mu=1}^{p}\xi_{i}^{\mu}\xi_{j}^{\mu}$,
where $\xi_{i}^{\mu}$ is a random variable which follows a bimodal
distribution.

One important aspect is that, in the corresponding mean field
approach using such $J_{ij}$ values, it is possible to introduce a
parameter which allows to control the level of frustration in the
problem \cite{Magal2}. The first interesting result is that the
Kondo solution is robust in the large $J_{K}$ limit, no matter what
is the level of frustration. For weak frustration and small $J_{K}$,
below a certain temperature, it appears a SG solution. When the
temperature is further decreased, the SG solution is replaced by
Mattis states which have the same thermodynamics as a FE phase
\cite{Amit}. This result suggests that the situation found in
$CeNi_{1-x}Cu_{x}$ alloys would be an example of weak frustration.
Nevertheless, there is an important difference between this model
and the approach of the Gaussian distributed $J_{ij}$. For the kind
of disorder given by this generalized Mattis model, there is no
mixed phase solution for the order parameters, but on the contrary,
there is a first order phase transition between the SG and Mattis
states; such solutions can obviously coexist, but one of
them is always metastable. In conclusion, our previous Mattis-like model gives the SG phase above the FE phase, 
but it cannot yield a real SG+FE mixed Phase \cite{Magal2}.

Thus, in order to improve the preceding description and to have,
therefore, a better agreement with experiment, we introduce here, in
our previously used KIL model, a new kind of site disordered
coupling $J_{ij}$, originally introduced by van Hemmen (vH) to study
the Spin Glass in the classical Ising model \cite{van Hemmen}. The
phase diagram obtained from such a classical model displays not only
SG, FE+SG and FE phases, but also they can appear in that order when
temperature is decreased. In this particular case, the SG+FE phase is characterized by both non 
zero magnetization and  SG order parameters. Recently, a work \cite{Ricardo1} has
studied a mean field solution of a quantum version of the vH model
with an applied transverse field $\Gamma$ and it shows that some
aspects of its classical counterpart can still be preserved in the
quantum vH model, and in particular the SG+FE phase. However, spin
flipping introduced by the presence of $\Gamma$ in the quantum vH
model can modify the phase diagram, suppressing for instance the
presence of SG+FE phase \cite{Ricardo1}. However, it is well known
that an additional transverse field in the KIL model with a Gaussian
random bond coupling between the localized Ising spins operators can
produce important consequences as, for instance, a QCP \cite{Alba2}.

In the present work, we will, therefore, study the KIL model with
both the vH type of disorder for the intersite exchange interaction
$J_{ij}$ and a transverse field $\Gamma$ which allows also to
investigate the possible consequences for the phase diagram with the
spin flipping. There is also another very important aspect related
to the vH type of disorder introduced in the present work:
in the previous approaches using the Gaussian random bond SK-type
$J_{ij}$ \cite{Alba1,Magal1,MagalAF1,MagalAF2,Alba2} or the site
disorder type given by the product of random variables
$\xi_{i}^{\mu}$, the disorder is treated using the so called replica
symmetry solution for the SG order parameters \cite{ReviewBinder}.
This solution is well known to have a serious flaw, because it is
locally unstable below the freezing temperature  \cite{Almeida}.
Certainly, that problem could be overcomed by the use of replica
symmetry breaking schemes \cite{Parisi}. However, this kind of
scheme increases the number of order parameters in such a way that
the search for order parameter solutions in the KIL model becomes
extremely complicated. Nevertheless, that is not the only problem
with the use of replicas to treat the disorder in the KIL model.
There are also indications that the presence of one or other
magnetic solutions could be dependent on the particular kind of
replica symmetry breaking schemes  \cite{ScesHouston}. By contrast,
that is not the case for the disordered $J_{ij}$ given in the vH
model (see following equation (\ref{eq3})). The disorder can be
treated without the use of replica technique as demonstrated in the
classical and quantum vH models \cite{van Hemmen,Ricardo1}. Thus,
the present use of the van Hemmen description of the disorder in the
KIL model improves considerably the description of the Kondo-Spin
glass-Ferromagnetism competition in disordered Kondo systems. 
It is important to remark that the present work 
is typically a mean field theory as in 
Refs. \cite{Alba1,Magal1,MagalAF1,MagalAF2,Alba2}. 
In particular, the Static and saddle point approximations are used here. 
The use of the first approximation can be justified since our goal 
is mainly to describe phase boundaries as 
discussed in Ref. \cite{Alba2}.
The saddle point method is in fact exact here, as a 
consequence of the long range nature of the vH coupling.

This paper is structured as follows. In the next section, we
introduce the model and calculate the corresponding thermodynamics.
The following section is dedicated to discuss the numerical
solutions of the saddle point equations for the order parameters and
to derive the phase diagram. Finally, the last section is reserved
to the conclusions.

\section{General Formulation}

The starting Hamiltonian in the KIL model is given by:
\begin{equation}
 \begin{split}
H={\displaystyle\sum_{ij,s}}t_{ij}\hat{n}_{is}^{d}+\epsilon_{0}
{\displaystyle\sum_{i,s}}\hat{n}_{is}^{f}
+J_{K}{\displaystyle\sum_{i}}[\hat{S}_{fi}^{+}\hat{s}_{di}^{-}+\hat{S}_{fi}^{-}\hat{s}_{di}^{+}]
\\
-{\displaystyle\sum_{i,j}}J_{ij}\hat{S}_{fi}^{z}S_{fi}^{z}-
2\Gamma\sum_{i}\hat{S}_{fi}^{x}.
\label{eq2}
\end{split}
\end{equation}
In  Eq. (\ref{eq2}), $\hat
S_{fi}^{z}=\frac{1}{2}[\hat{n}_{i\uparrow}^{f}-\hat{n}_{i\downarrow}^{f}]$,
$\hat S_{fi}^{+}=f_{i\uparrow}^{\dagger}f_{i\downarrow}$,
$\hat{S}_{fi}^{-}=(\hat{S}_{fi}^{+})^{\dagger}$,
$\hat S_{fi}^{x}=f_{i\uparrow}^{\dagger}f_{i\downarrow}+f_{i\downarrow}^{\dagger}f_{i\uparrow}$,
$\hat s_{di}^{+}=d_{i\uparrow}^{\dagger}d_{i\downarrow}$,
$\hat{s}_{di}^{-}=(\hat{s}_{si}^{+})^{\dagger}$,
$\hat{n}_{is}^{f}=f_{is}^{\dagger}f_{is}$,
$\hat{n}_{is}^{d}=d_{is}^{\dagger}d_{is}$ where
$f_{is}^{\dagger}~(f_{is})$ and $d_{is}^{\dagger}~(d_{is})$  are
fermionic creation (destruction) operators of $f$ and $d$ electrons,
respectively. The spin projections are indicated by $s=\uparrow$ or
$\downarrow$.

The random coupling $J_{ij}$ in Eq. (\ref{eq2}) is given as in the
vH model by:
\begin{equation}
J_{ij}=\frac{J}{N}(\xi_{i}\eta_{j}+\eta_{i}\xi_{j})+\frac{J_{0}}{N}
\label{eq3}
\end{equation}
where $\xi_{i}$ and $\eta_{i}$ in Eq. (\ref{eq3}) are random
variables which follow the bimodal distribution:
\begin{equation}
 P(x)=\frac{1}{2}[\delta(x-1)+\delta(x+1)].
\label{eq4}
\end{equation}
In Eq. (\ref{eq4}), $\delta(x)$ is the Dirac delta function.
As discussed in the previous section,
the coupling $J_{ij}$ given in Eq. (\ref{eq3}) is an infinite long range
coupling which 
gives 
exact solutions in the thermodynamical limit for the 
saddle point approximation used below.

The partition function is expressed within functional formalism
using anticommuting Grassmann variables $\varphi_{is}(\tau)$
and $\psi_{is}(\tau)$ associated to the $f$ and $d$ electrons, respectively
as \cite{Alba1,Alba2}:
\begin{equation}
Z  = \int D(\psi^{\ast}\psi) D(\varphi^{\ast}\varphi)
\exp\left[A_{VH} + A_K
+ A_0 \right].
\label{e8}
\end{equation}
In the static approximation (SA) \cite{Moore}
the actions in Eq (\ref{e8}) are given as:
\begin{equation}
 \begin{split}
A_0  =\sum_{\omega} \sum_{i,j}
 \left[(\underline{\psi})_{i}^{\dagger}(\omega)
(i\omega -\beta \varepsilon_{0}+\beta\Gamma \underline{\sigma}_x)
\delta_{ij} \underline{\psi}_{i}(\omega)\right.
\\
\left. +
\underline{\varphi}_{i}^{\dagger} [(i\omega+\mu_{d})\delta_{ij} -
\beta t_{ij}]\underline{\varphi}_{j}
(\omega) \right]
\label{A0}
\end{split}
\end{equation}
where, in the first term of Eq. (\ref{A0}), the chemical potential
$\mu_{f}$ has been absorbed in $\varepsilon_{0}$.
\begin{equation}
 \begin{split}
A^{stat}_K \approx \frac{J_K}{N} \sum_{is} \sum_{\omega}
\left[ \varphi_{i-s}^{\ast}(\omega)
\psi_{i-s}(\omega)\right]
\sum_{js}\sum_{\omega^{'}}\left[ \psi_{js}^{\ast}(\omega^{'})
\varphi_{js}(\omega^{'})\right],
\end{split}
\end{equation}
\begin{equation}
A^{stat}_{VH} ={\displaystyle \sum_{ij}} J_{ij} S_{fi}^{z}S_{fj}^{z}
\label{e9}
\end{equation}
with
\begin{eqnarray}
S_{fi}^{z}=\frac{1}{2}{\displaystyle \sum_{\omega}}\underline{\psi}_{i }^{\dagger}(\omega)\underline{\sigma}^{z}\psi_{i }(\omega).
\label{eq10}
\end{eqnarray}

The action $A_{K}^{stat}$ is given in the mean field approximation (see Ref \cite{Alba1}).
In the remaining components of the action $A_{0}$ and $A_{VH}^{stat}$, spinors are used :
\begin{eqnarray}
\begin{array}{ccc}
 \underline{\varphi}_{i}(\omega)=\left( \begin{array}{c} \varphi_{i\uparrow}(\omega) \\ \varphi_{i\downarrow}(\omega)\end{array}\right),  & &
\underline{\psi}_{i}(\omega)=\left(\begin{array}{c} \psi_{i\uparrow}(\omega) \\ \psi_{i\downarrow}(\omega)\end{array}\right)
\end{array}
\label{ee369b}
\end{eqnarray}
and the Pauli matrices are given as usual by:
\begin{eqnarray}
 \begin{array}{ccccc}
 \underline{\sigma}_{x}=\left(\begin{array}{cc} 0 & 1 \\ 1 & 0
 \end{array}
 \right)& &\underline{\sigma}_{y}=\left(\begin{array}{cc} 0 & -i \\ i & 0
 \end{array}
 \right) &\underline{\sigma}_{z}=\left(\begin{array}{cc} 1 & 0 \\ 0 & -1
 \end{array}
 \right).\end{array}
 \label{ee369c}
 \end{eqnarray}

We follow a procedure close to that introduced in Ref. \cite{Alba1}.
Therefore, the Kondo order parameter
$\lambda_{\sigma}\approx\lambda=\frac{1}{N}{\displaystyle\sum_{j,\omega}}\left\langle
\psi_{j\sigma}^{*}(\omega)\varphi_{i\sigma}(\omega)\right\rangle$
can be introduced in the partition function $Z$ defined by Eqs.
(\ref{e8})-(\ref{eq10}). Then, the $\varphi$ fields are integrated
and we obtain the following result:
\begin{equation}
Z/Z^{0}_{d}=\exp(-2N\beta \lambda \lambda^{*})Z_{eff}
\label{eq5}
\end{equation}
where $Z^{0}_{d}$ is the partition function  of free $d$ electrons and
\begin{equation}
 \begin{split}
Z_{eff}={\displaystyle \int D(\psi^{*}\psi)} \exp\left( A^{stat}_{VH} 
{+\displaystyle\sum_{\omega\sigma}}{\displaystyle\sum_{i,j}}\underline{\psi}_{i\sigma}^{*}
(\omega)\underline{g}_{ij}^{-1}(\omega)
\underline{\psi}_{j\sigma}(\omega)
\right)
\label{eq6}
 \end{split}
\end{equation}
with
\begin{equation}\begin{split}
\underline{g}_{ij}^{-1}(\omega)=[(i\omega-\beta\epsilon_{0})\underline{I}+\beta\Gamma\underline{\sigma_{z}}]\delta_{ij}
-
\frac{\beta^{2}J_{k}^{2}\lambda^{2}}{(i\omega+ \mu_{d})\delta_{ij}-\beta t_{ij}}\underline{I}.
\label{eq7}
\end{split}\end{equation}
In Eq.(\ref{eq7}), we use the notation $|\lambda^{2}|\equiv\lambda^{2}$ and $\underline{I}$ means the unitary matrix.

Introducing $J_{ij}$ given by Eq. (\ref{eq3}), the action
$A^{stat}_{vH}$ becomes composed of two terms: one randomic and the
other one ferromagnetic. They can be rearranged to introduce SG and
FE order parameters in $Z_{eff}$. The details of such calculations
are shown in the Appendix.

The free energy is, therefore, given by:
\begin{equation}
\beta F= 2\beta J_{K} \lambda^{2}- \lim_{N\rightarrow\infty} \frac{1}{N}\ln Z_{eff}.
\label{eq100}
\end{equation}

Using the saddle point solution for $Z_{eff}$ (see Eqs.
((\ref{eq11})-(\ref{eq19}))), the free energy in Eq. (\ref{eq100})
becomes:
\begin{equation}
\begin{split}
 \beta F=2\beta J_{K} \lambda^{2}+2\beta J
q_{1}q_{2} + \beta J_{0}m^{2}
- {\displaystyle\sum_{\omega}}\ln(\det
\underline{G}_{ij}^{-1}(\omega|h_{j})).
\label{eq25}
\end{split}
\end{equation}
The Green function  $\underline{G}_{ij}^{-1}(\omega|h_{j})$
is given in Eqs. (\ref{eq20})-(\ref{eq15}).
In order to proceed to the calculations, we use in the last term of  Eq. (\ref{eq25})
the approximation introduced in Ref. \cite{Alba1}
which decouples the
random magnetic field $h_{j}$ from the
Kondo lattice.
Thus, we obtain:
\begin{equation}
\ln \det\left(
\underline{G}_{ij}^{-1}(\omega|h_{j})\right)\approx\frac{1}{N}
{\displaystyle\sum_{j}}\ln\left[\det\underline{\Gamma}_{\mu
\nu}^{-1}(\omega|h_{j})\right]
\label{eq23}
\end{equation}
with
\begin{equation}
 \begin{split}
\Gamma_{\mu\nu}^{-1}(\omega|h_{j})=\left[ (i\omega - \beta
\epsilon_{0})\underline{I}-\underline{\sigma_{z}} h_{j}
+\beta\Gamma\underline{\sigma_{x}} \right]\delta_{\mu \nu}
\\
-\beta^{2} J_{k}^{2}\lambda^{2}\frac{1}{N}
{\displaystyle\sum_{\vec{k}}} \frac{ e^{i\vec{k}\vec{R}_{\mu \nu}}
}{(i \omega+\mu_{d})-\beta\epsilon_{k}}\underline{I}.
\label{eq24}
 \end{split}
\end{equation}

Now, in the  last term of Eq. (\ref{eq25}), we can use
self-averaging property $\frac{1}{N}\sum_{j}
f(\eta_{j};\xi_{j})=\left\langle \left\langle f(\eta; \xi)
\right\rangle \right\rangle_{\xi\eta}$.
Therefore:
\begin{equation}
\begin{split}
\frac{1}{N}{\displaystyle\sum_{j}} \ln\left[
\frac{1}{N}{\displaystyle\sum_{\vec{k}}}(\sum_{\omega}\Gamma_{\sigma}(\vec{k},h_{j}))
\right]
 = \left\langle \left\langle\ln\left[
\frac{1}{N}{\displaystyle\sum_{\vec{k}}}(\sum_{\omega}\Gamma_{\sigma}(\vec{k},h))\right]
\right\rangle \right\rangle _{\xi\eta}
\label{eq29}
\end{split}
\end{equation}
where
\begin{equation}
\left\langle \left\langle
f(\xi,\eta)
\right\rangle \right\rangle _{\xi \eta}= {\displaystyle \int d\xi d\eta P(\xi, \eta) f(\xi,\eta)}
\label{eq30}
\end{equation}
Then, by assuming that $\mu_{d}=0$ and $\varepsilon_{0}=0$, the free
energy can be found as
\begin{equation}
 \begin{split}
\beta F=  2\beta J_{k}\lambda^{2} + 2\beta Jq_{1}q_{2} + \beta J_{0}m^{2}
\\
-
\left\langle \left\langle
\frac{1}{\beta D}
{\displaystyle \int_{-\beta D}^{+ \beta D}} dx \ln
\left[
\cosh\left(\frac{x+H}{2} \right)
\right.\right.\right.
\\
\left.\left.\left.
+\cosh \sqrt{\frac{1}{4}(x-H)^2+\beta^{2}J_{k}^{2}\lambda^{2}}
\right]\right\rangle \right\rangle _{\xi\eta}
\label{eq33}
 \end{split}
\end{equation}
with
\begin{equation}
H = \beta\sqrt{\left[2J(\eta q_{2}+\xi q_{1})+ 2J_{0}m
\right]^{2}+\Gamma^{2} }.
\label{eq211}
\end{equation}

In Eq. (\ref{eq33}), the sums over the Matsubara frequencies and
over $\vec{k}$ have been done in a way similar to Ref. \cite{Alba1}. 
We have also used here the usual approximation of a constant density of states
for
the $d$ electrons, $\rho=\frac{1}{2D}$ for $-D<\epsilon<D$. 
 The use of this density of states allows a direct comparison   
of phase diagrams obtained in this work with previous ones given
in Refs. \cite{Alba1,Magal1,Magal2}.
Finally,
assuming that the probability distribution $P(\xi,
\eta)=P(\xi)P(\eta)$, we can compute $\left\langle\left\langle...
\right\rangle \right\rangle_{\xi \eta}$ in Eq. (\ref{eq33}) using
Eqs. (\ref{eq4}) and (\ref{eq30}). 

\begin{figure}[t]
 \includegraphics[height=\columnwidth,angle=-90]{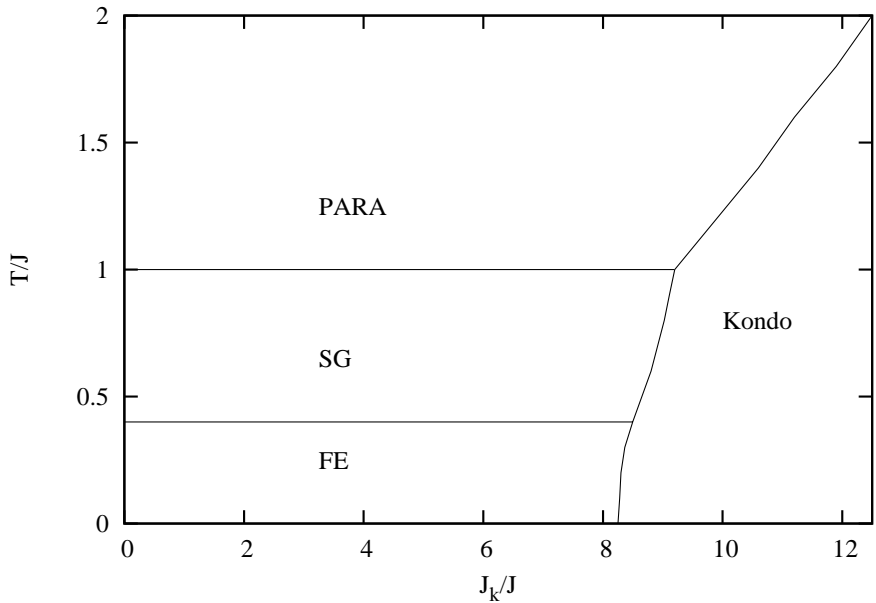}
\caption{Phase diagram $T/J$ {\it versus} $J_K/J$ for $J_0/J=1.6$ and $\Gamma/J=0$.}
\label{fig1}
\end{figure}

\section{Numerical results}

The coupled saddle point equations for $q_{1}$, $q_{2}$, $m$ and
$\lambda$ can be obtained directly from Eqs.
(\ref{eq33})-(\ref{eq211}). The numerical solutions for such order
parameters allow us to obtain the following phases: (i)
paramagnetism (PARA) given by $q_{1}=q_{2}= 0$, $m=0$ and
$\lambda=0$; (ii) the SG phase given by $q_{1}=q_{2}\neq 0$, $m=0$
and $\lambda=0$; (iii) the mixed phase (SG+FE) given by
$q_{1}=q_{2}\neq 0$, $m\neq 0$ and $\lambda=0$; (iv) ferromagnetism
(FE) given by $q_{1}=q_{2}= 0$, $m\neq 0$ and $\lambda=0$; (v) Kondo
state where only $\lambda$ is different from zero. For
numerical results, $D/J=12$ is used.

\begin{figure}[t]
\includegraphics[height=\columnwidth,angle=-90]{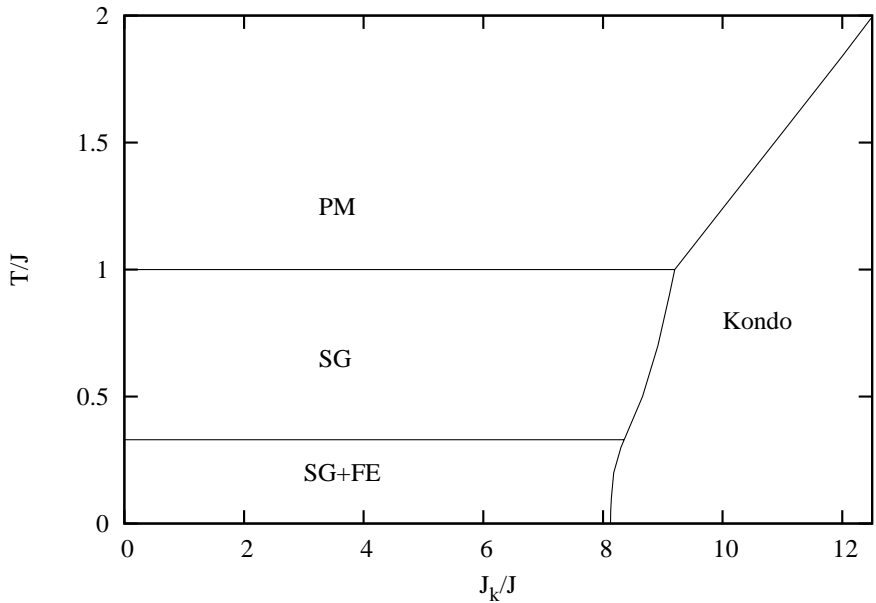}
\caption{Phase diagram $T/J$ {\it versus} $J_K/J$ for $J_0/J=1.3$ and $\Gamma/J=0$.}
\label{fig2}
\end{figure}

\begin{figure}[ht]
 \includegraphics[height=\columnwidth,angle=-90]{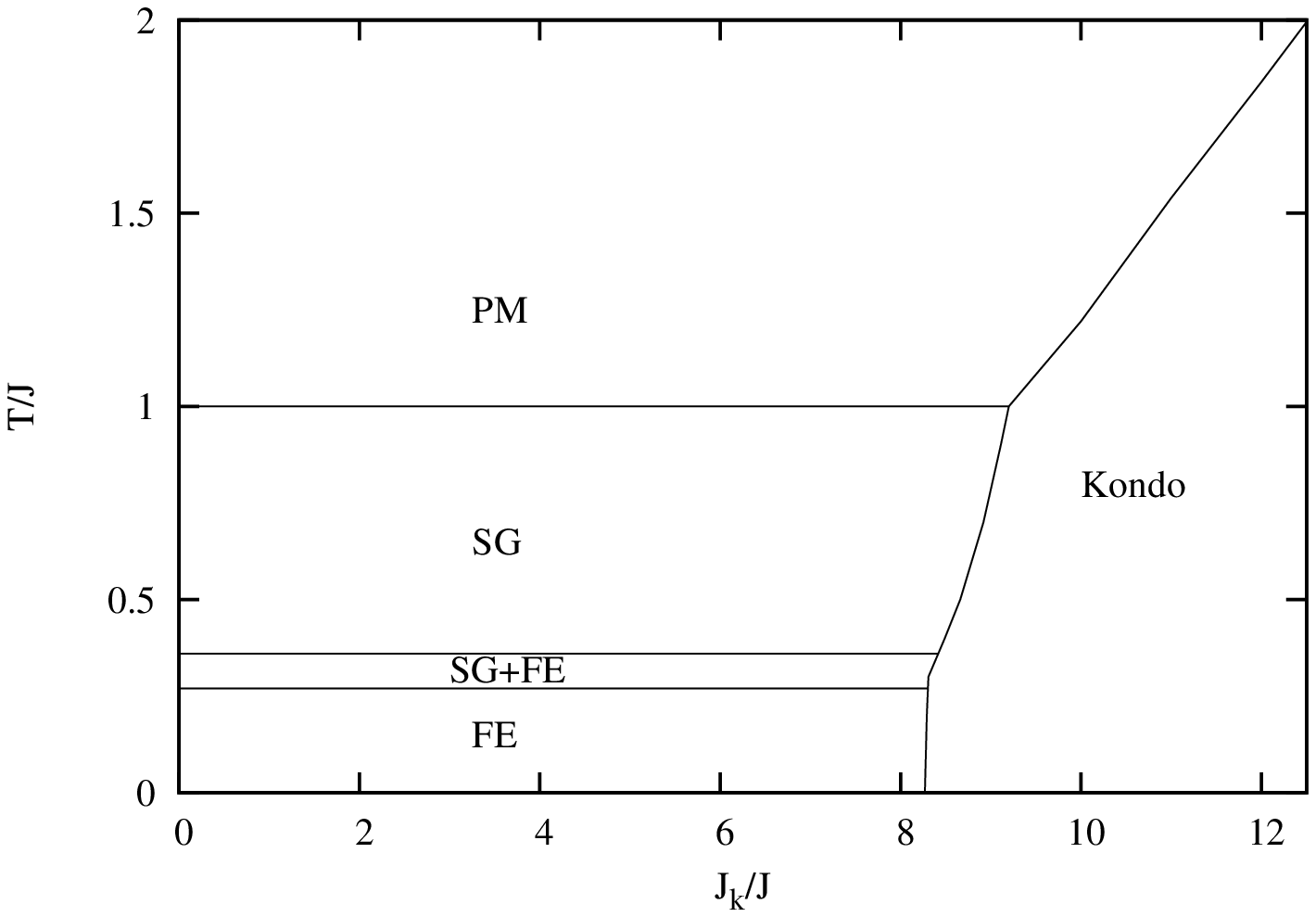}
\caption{Phase diagram $T/J$ {\it versus} $J_K/J$ for $J_0/J=1.4$ and $\Gamma/J=0$ .}
\label{fig3}
\end{figure}

Phase diagrams giving temperature $T$ {\it versus} $J_K$ (in units
of $J$) can be built for several values of $J_{0}/J$ and $\Gamma/J$.
In Figure (\ref{fig1}), such a phase diagram is displayed for
$J_0/J=1.6$ and $\Gamma/J=0$. For this case, in the large $J_K$
region there is only one solution which corresponds to the Kondo
state. When $J_K$ decreases, the Kondo solution disappears.
Actually, it is substituted by the magnetic solutions PARA, SG and
FE which appear in that order when $T$ is lowered. In Figure
(\ref{fig2}), we take $J_0/J=1.3$ and $\Gamma=0$. This decrease of
$J_0/J$ from 1.6 to 1.3 does not affect the Kondo state, but changes
a lot the magnetic solutions. In Figure (\ref{fig2}), the solution
FE  is replaced by the mixed phase SG+FE, while the size of the
region where the SG solution exists remains almost the same as in
Figure (\ref{fig1}). In Figure (\ref{fig3}), the transverse field
$\Gamma$ is maintained equal to 0 and we take an intermediate value
$J_0/J=1.4$. As in the two previous cases, the Kondo state is not
really affected in the large $J_K$ region, but the region of the
magnetic solutions in the phase diagram is again modified. Besides
the existence of SG and SG+FE solutions, when the temperature is
decreased, there is also an additional FE solution at much lower
temperatures. In  other words,  in a small range of $J_0/J$
($1.3\leq J_0/J \leq 1.6$), the phase diagrams present several
scenarios concerning the existence of magnetic solutions. In
contrast, the Kondo state is robust to such changes of $J_0/J$.

\begin{figure}[ht]
 \includegraphics[height=\columnwidth,angle=-90]{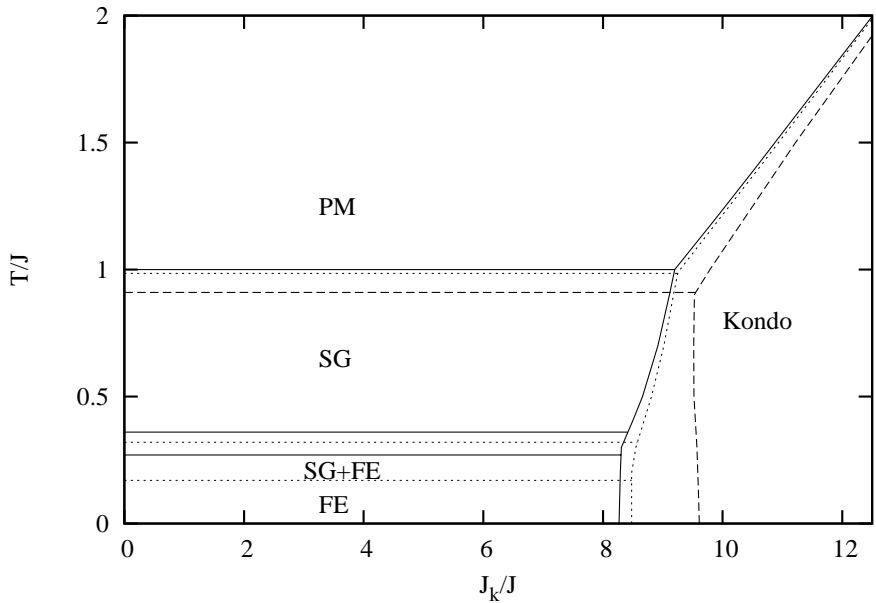}
\caption{Phase diagrams $T/J$ {\it versus} $J_K/J$ for $J_0/J=1.4$
and three values of $\Gamma/J:$ 0, 0.4 and 1.0. The dashed, dotted and full lines are
results for $\Gamma/J=1.0$, $\Gamma/J=0.4$ and $\Gamma/J=0.0$, respectively. 
The critical lines for
$\Gamma/J=0$ occur at higher temperatures than those ones for
$\Gamma/J=0.4$ and $\Gamma/J=1.0$. In particular, for $\Gamma/J=1.0$ there is no more SG+FE and FE solutions. 
} \label{fig4}
\end{figure}

Furthermore, a new situation is obtained when the transverse field
is turned on, as can be seen in Figure (\ref{fig4}). For instance,
for $\Gamma=1.0$, the Kondo solution is obtained for a value of
$J_K/J$ a little larger than that found previously for $\Gamma=0.4$
or $\Gamma=0$ and simultaneously, the range of $J_K/J$ where the
magnetic solutions are found is increased. Moreover, for such a
decrease of $\Gamma$ from 1 to 0, the transition temperatures
between the magnetic phases are clearly depressed. But the most
important feature observed with the increase of $\Gamma$ concerns
the magnetic solutions, because the SG+FE and FE phases disappear
completely and it remains only the SG phase for a sufficiently large
$\Gamma$ value.


\section{Conclusions}

In the present work, the KIL model has been studied with assuming
that the inter-site spin coupling $J_{ij}$ between localized spins is a
random coupling given by the van Hemmen model as given in Eq.
(\ref{eq3}). It has also been added to the model a transverse field $\Gamma$
which mimics a Heisenberg spin-flipping term.

The results are shown in Figures (\ref{fig1})-(\ref{fig4}). For
$\Gamma=0$, they basically display two regimes when the strength
$J_K$ of the Kondo interaction is varied in units of the 
component $J$ of the coupling $J_{ij}$ (see Eq. (\ref{eq3})). In the
first regime obtained for large $J_K$ values, there is only the
Kondo phase. In contrast, the second regime with only the magnetic
solutions SG, SG+FE and FE exists when $J_{K}$ is decreased. One
important point is the order in which the magnetic phases are found
when the temperature is decreased. For instance, the SG phase is
found at higher temperature. Then, it can  appear a SG+FE phase. The
pure FE phase is found only at the lowest temperatures. It is also
important to notice that the existence of the different solutions
SG, SG + FE or FE depends on the strength of the ferromagnetic
component $J_0$ (given in units of $J$) of the coupling $J_{ij}$, as
can be seen in the Figures (\ref{fig1})-(\ref{fig3}). When $\Gamma$
is different from zero, the two regimes discussed previously are
affected. While the Kondo solution needs larger values of $J_K$ to
be found, the magnetic solutions found at lower temperatures
disappear rapidly when $\Gamma$ is increased.

It should be emphasized that the present approach using the $J_{ij}$
coupling given by the vH model yields two important improvements
with respect to previous approaches. The first one concerns the use
of the replica method which is not necessary here to generate the
thermodynamics. This is an important improvement with respect to the
previous approaches using the bond disorder given by the SK-like
Gaussian random $J_{ij}$ in the KIL model
\cite{Alba1,Alba2,MagalAF1,MagalAF2,Magal1} or using the previous
Mattis-like approach \cite{Magal2}. For instance, the presence of
magnetic solutions in these approaches is quite dependent on which
particular scheme of replica solution is used, as explained in the
discussion of Ref. \cite{ScesHouston}.

As our present results suggest, the second improvement 
concerns the
particular kind of site disorder given by the vH model introduced in
the KIL model with a certain range of $J_0$, which allows to obtain
magnetic solutions SG, SG+FE and  FE phases when the temperature is
decreased. In that sense, the weakness of the approach proposed in
Reference \cite{Magal2} is overcome and we are able to introduce
here a mixed phase SG+FE.

 Thus, our present calculation using the van Hemmen site
disorder can describe Cerium disordered physical systems such as
$CeNi_{1-x}Cu_{x}$ or $CePd_{1-x}Rh_{x}$ alloys. In particular,
Figures 3 and 4 can describe the phase diagram of
$CeNi_{1-x}Cu_{x}$ with $J_K$ increasing with an increasing
Nickel concentration, by explaining the Kondo behavior observed
for $x$ close to $1$ and by proposing a good interpretation of the complicate magnetic behavior
observed for smaller $x$ values. 
These are indications that the use of the van Hemmen site disorder could
be useful to describe physical systems such as $CeNi_{1-x}Cu_{x}$ or
$CePd_{1-x}Rh_{x}$ alloys, although the low temperature phase is in
these alloys a Kondo-cluster-glass followed by a disordered
ferromagnetic one. However, it is important to notice that canonical
spins have been used in the present work. This description is
obviously not sufficient to capture the complexity of the cluster glass
state. However, earlier results for a mean field formulation of the
cluster glass indicate that there are no essential differences
between canonical spins and clusters of spins, as far as the phase
boundaries are concerned \cite{Sokoulis}. One can, therefore, expect
that most of the previous discussion concerning
the sequence of magnetic orders as a function of $J_K$
can
be preserved even if the problem is formulated in terms of clusters
of spins instead of canonical spins as it is done in the present
work.

On the other hand, we are presently working on a theoretical
description of the Kondo-Cluster-Glass, by solving exactly the
problem in a small cluster with $n_s$ atoms interacting between them
by a disorder spin glass-like interaction. We have already solved
the problem with only $n_s=3$ and a disorder intercluster bonding given by the
Sherrington-Kirkpatrick interaction \cite{ScesBuzios}. We think that
the van Hemmen approach is easier to treat and we are presently
working on clusters with a larger number $n_s$, in order to have
finally a more local description of the Kondo-Cluster-Glass observed
in some disordered Kondo Cerium systems.

In conclusion, we have to remark that our van Hemmen-Kondo
description yields considerable improvements with respect to
previous theoretical models in the two following points, the non
consideration of the replica method and the problem of the mixed SG+FE
phase. The validity of the van Hemmen model, which does not use the replica trick method, has been
discussed in detail and it has been shown that this model is
perfectly able to describe the spin glass experiments and that it is
simpler than the other models for a mathematical treatment
\cite{Choy,van Hemmen1}. On the other side, our van Hemmen-Kondo model gives with decreasing temperature a SG phase, a SG+FE one and finally a ferromagnetic phase and the intermediate SG+FE phase is a real mixed phase with together non zero SG and FE order parameters. This model gives a good account for the experimental phase diagrams of disordered Cerium systems, such as $CeNi_{1-x}Cu_{x}$ alloys,
and can be used to have a more local description of the Kondo-Cluster-Glass phase.

\section*{Acknowledgments}
B. Coqblin acknowledges the European Cost P16 Action for financial
support.
S.G. Magalhaes and F.M. Zimmer acknowledge the CNPq for financial support.

\appendix
\section*{Appendix}
\setcounter{section}{1}

In this appendix, we present in details the procedure which allows
to introduce the SG and FE order parameters in the problem. First,
the random component of $J_{ij}$ given in Eq. (\ref{eq3}) can be
rewritten as:
\begin{equation}
\begin{split}
\frac{\beta J}{N}{\displaystyle\sum_{i\neq j}}(\eta_{i}\xi_{j}+\xi_{i}\eta_{j})S_{i}^{z}S_{j}^{z}=
\\
\frac{\beta J}{N}
   \left[ {\displaystyle\sum_{j=1}^{N}}(\eta_{j}+\xi_{j})S_{i}^{z}\right]^{2}
   -\frac{\beta J}{N}
  \left[  {\displaystyle\sum_{j=1}^{N}}\eta_{i}S_{i}^{z}\right] ^2
\\
-\frac{\beta J}{N} \left[
{\displaystyle\sum_{j=1}^{N}}\xi_{i}S_{i}^{z}\right] ^2
-\frac{2\beta
J}{N}{\displaystyle\sum_{j=1}^{N}}(\eta_{i}S_{i}^{z})(\xi_{i}S_{i}^{z})
\label{eq9}
\end{split}
\end{equation}
while the ferromagnetic one is
\begin{equation}
 \frac{\beta J_{0}}{N} {\displaystyle\sum_{i\neq j}}S_{i}^{z}S_{j}^{z}=\frac{\beta J_{0}}{N}
\left[ {\displaystyle\sum_{i}}S_{i}^{z}\right] ^2-\frac { J_{0}}{N} {\displaystyle\sum_{i}}\left( S_{i}^{z}\right)^2.
\label{eq101}
\end{equation}
The last terms in Eqs (\ref{eq9}) and (\ref{eq101}) vanish in the thermodynamic limit.

The Hubbard-Stratonovich transformation can be used to linearize the
action $A^{stat}_{vH}$.
Thus $Z_{eff}$ in Eq. (\ref{eq6}) becomes:
\begin{equation}
\begin{split}
Z_{eff}= \left( \frac{N}{2\pi}\right)^{2} {\displaystyle \int_{-
\infty}^{+ \infty}} d \bar{q}_{1}{\displaystyle \int_{- \infty}^{+
\infty}} d \bar{q}_{2} {\displaystyle \int_{- \infty}^{+ \infty}} d
\bar{q}_{3}
\\
\times{\displaystyle \int_{- \infty}^{+ \infty}} dm
\exp \left( -\frac{N}{2} (\bar{q}_{1}^{2}+ \bar
{q}_{2}^{2}+ \bar{q}_{3}^{2})- \frac{N \bar{m}^{2}}{2}
  +  \ln
\Lambda (\bar{q}_{1},\bar{q}_{2},\bar{q}_{3}, \bar{m}) \right)
\label{eq11}
\end{split}
\end{equation}
where the function $\Lambda (\bar{q}_{1},\bar{q}_{2}, \bar{q}_{3},
\bar{m})$ in Eq.(\ref{eq11}) is:
\begin{equation}
\begin{split}
\Lambda(q_{1}, q_{2},q_{3},m)= {\displaystyle \int
D(\psi^{*}\psi)}
 \exp \left(
{\displaystyle\sum_{i,\sigma}}{\displaystyle\sum_{\omega}}\underline{\psi}_{i
\sigma}^{*} \underline{G}_{ij}^{-1}(\omega|h_{j})\underline{\psi}_{j
\sigma}(\omega)\right) \label{eq19}
\end{split}\end{equation}
with:
\begin{equation}
\begin{split}
 \underline{G}_{ij}(\omega|h_{j})=
\left[
(i\omega +\beta \epsilon_{0})\underline{I} -\underline{\sigma_{z}} h_{j}+\beta\Gamma\underline{\sigma_{x}}
\right] \delta_{ij}
 -\frac{\beta^{2}J_{k}^{2}\lambda^{2}}{i (\omega+\mu_{d})\delta_{ij}-\beta t_{ij}}\underline{I}.
\label{eq20}
\end{split}
\end{equation}
The random field in Eq.(\ref{eq20}) is
\begin{equation}
h_{j}
= \sqrt{2\beta J} \left(i\eta_{j} \bar{q}_{1}+ i \xi_{j}
\bar{q}_{2}+(\eta_{j}+\xi_{j})\bar{q}_{3}\right)+ \sqrt{2\beta J_{0}}\bar{m}
\label{eq21}
\end{equation}
The saddle point solution of Eq. (\ref{eq11}) gives:
\begin{equation}
\bar{q}_{1} = i \sqrt{\beta J} \frac{1}{N}\sum_{j}\left\langle
\xi_{j} S^{z}_{j} \right\rangle
= i \sqrt{2\beta J}q_{1}
\label{eq16}
\end{equation}
\begin{equation}
\bar { q_{2}} = i \sqrt{\beta J}
\frac{1}{N}\sum_{j}\left\langle \eta_{j} S^{z}_{j}\right\rangle
= i \sqrt{2\beta J}q_{2}
\label{eq17}
\end{equation}
\begin{equation}
\bar {q}_{3}
=  \sqrt{2\beta J} \left( q_{1}+q_{2}\right)
\label{eq18}
\end{equation}
and
\begin{equation}
\bar{m}=\sqrt{2\beta J_{0}}\frac{1}{N}\sum_{j} \left\langle
S^{z}_{j}
\right\rangle
= \sqrt{2 \beta J_{0}} m
\label{eq15}
\end{equation}
The symbol $\left \langle ... \right\rangle$ is the thermodynamical
average and $i^2 = -1$ in Eqs. (\ref{eq16})-(\ref{eq15}). The
integral over the Grasmann fields can be performed in
Eq.(\ref{eq19}), leading to:
\begin{equation}
\Lambda(q_{1},q_{2},m)= \exp
\left(
{\displaystyle\sum_{\omega}}\ln(\det \underline{G}_{ij}^{-1}(\omega|h_{j}))
\right).
\label{eq22}
\end{equation}
%


\end{document}